\documentstyle[epsf]{kluwer}
\input psfig.sty
\def\gapprox{\;\rlap{\lower 2.5pt            
 \hbox{$\sim$}}\raise 1.5pt\hbox{$>$}\;}       
\def\lapprox{\;\rlap{\lower 2.5pt            
 \hbox{$\sim$}}\raise 1.5pt\hbox{$<$}\;} 
\def\N{\,{\rm I\kern-.20em N}}
\begin{document}

\begin{article}
\begin{opening}
\title{Survey on solar X-ray flares and associated coherent radio emissions}
\author{Arnold O. \surname{Benz} $^1$ }
\author{Paolo C. \surname{Grigis}$^1$}
\author{Andr\'e \surname{Csillaghy}$^{1,2}$}
\author{Pascal \surname{Saint-Hilaire}$^{1,3}$}

\institute{$^1$ Institute of Astronomy, ETH Z\"urich, CH-8092 Zurich, Switzerland\\
$^2$ University of Applied Sciences, CH-5210 Windisch, Switzerland\\
$^3$ Paul Scherrer Institute, CH-5232 Villigen, Switzerland\\}
\runningtitle{X-ray flares and coherent radio emissions}
\runningauthor{Arnold O. Benz et al.}

\date{Received:.......  ; accepted :....... }
\begin{abstract} 
The radio emission during 201 X-ray selected solar flares was surveyed from 100 MHz to 4 GHz with the Phoenix-2 spectrometer of ETH Zurich. The selection includes all RHESSI flares larger than C5.0 jointly observed from launch until June 30, 2003. Detailed association rates of radio emission during X-ray flares are reported. In the decimeter wavelength range, type III bursts and the genuinely decimetric emissions (pulsations, continua, and narrowband spikes) were found equally frequently. Both occur predominantly in the peak phase of hard X-ray (HXR) emission, but are less in tune with HXRs than the high-frequency continuum exceeding 4 GHz, attributed to gyrosynchrotron radiation. In 10\% of the HXR flares, an intense radiation of the above genuine decimetric types followed in the decay phase or later. Classic meter-wave type III bursts are associated in 33\% of all HXR flares, but only in 4\% they are the exclusive radio emission. Noise storms were the only radio emission in 5\% of the HXR flares, some of them with extended duration. Despite the spatial association (same active region), the noise storm variations are found to be only loosely correlated in time with the X-ray flux. In a surprising 17\% of the HXR flares, no coherent radio emission was found in the extremely broad band surveyed. The association but loose correlation between HXR and coherent radio emission is interpreted by multiple reconnection sites connected by common field lines.
\end{abstract} 
\end{opening}
\section{Introduction}
`Flares are highly individualistic', already noticed Smith \& Smith (1963) reporting on optical observations. They recapitulated the first century of flare observations and were impressed by the different morphology of each H$\alpha$ flare. Today's observers of the non-thermal aspects of the flare phenomenon, particularly of the hard X-rays (HXR) and decimeter radio emissions, have a similar experience. The pattern of the various emissions in the radio spectrogram combined with the X-rays seems to be equally disparate in each flare. 

In meter waves, the classification into 5 types has been sufficient to characterize, interpret and model the emissions. A comprehensive review of the meter-wave emissions may be found in a volume edited by McLean \& Labrum (1985). The situation is not clear, however, in decimeter waves where in principle a similar number of types exists but seems to represent still largely morphological classes in spectrograms rather than physical phenomena. The decimetric radio emission has been classified by G\"udel \& Benz (1988) and Isliker \& Benz (1994) into type III-like bursts, pulsations, diffuse continua, narrowband spikes, decimetric type IV bursts, and high-frequency continuum generally attributed to gyrosynchrotron emission. Except for the last one, these classes are rather broad and may well include emissions originating by different processes. Each includes several subclasses or various fine structures. The above 6 broad classes have basically been confirmed by other observers including Allart et al. (1990) and Bruggmann et al. (1990) at frequencies up to 8.5 GHz, and enriched by further subclasses (e.g. Jiricka, Karlicky, \& Meszarosova, 2002). The details of the various burst classes and subclasses are not reviewed here. The reader is referred to the original literature quoted and to the review by Bastian, Benz, \& Gary (1998). 

Individual radio burst types have been studied extensively for the association with HXRs. Kane (1981) notes that only 3\% of meter-wave type III bursts are correlated with HXR emission, but the correlation increases with higher start frequency of the bursts. Dennis et al. (1984) and Aschwanden et al. (1985) find occasional associations between reverse-drift type III bursts with HXR peaks. Aschwanden et al. (1995) took a large sample of HXR peaks and find associated type III bursts around 300 MHz in 31\%  of all cases. Some cases of correlation between the start or enhancement of a type I bursts (also called `noise storm') and soft X-rays were reported by Lantos et al. (1981), Raulin \& Klein (1994) and Willson, Kile, \& Rothberg (1997). HXR emission was found coincident with only 20\% of all decimetric pulsations by Aschwanden, Benz, \& Kane (1990). However, the more irregular subtype of drifting pulsations was reported as well associated in several recent case studies (Kliem, Karlick\'y, \& Benz, 2000; Kahn et al., 2002; Saint-Hilaire \& Benz, 2003). Finally, 95\%  of all decimetric spike events were found associated with HXR emission, but only 2\% of the HXR events were associated with spikes (G\"udel, Aschwanden \& Benz, 1991). 

General surveys on coherent radio emissions during X-ray flares are rare. Aschwanden et al. (1985) used observations by ISEE-3 in the two channels 26 - 43 keV and 5.8 - 6.9 keV. The radio data consisted of analog recordings on film and were limited to the 100 - 1000 MHz range. The prime list was selected from radio events. Simnett \& Benz (1986) did the opposite and selected strong HXR events ($>$ 1000 cts/s at $>$ 25 keV) from the HXRBS/SMM list. They found no radio emission in 15\% of the events, using data from the same analog spectrometer.

Still missing is a broadband survey of radio emissions during `X-ray selected' flares. X-ray emission is the widely used measure for flare energy and thus selects the more energetic flares. The key questions are: Which flares emit what type of radio emission and in which phase of the flare? What are the general patterns for impulsive HXR flares?

In this study, two large and homogeneous data bases of X-ray and radio observations are used to address the question of HXR-radio association. The RHESSI detector resolves well the 10 - 20 keV range where most of the flare photons are detected. It is the first time that the Phoenix-2 radio spectrometer was used to survey its entire possible band from 100 MHz to 4 GHz for an extended period. The band covers the entire frequency range of the plasma frequency and electron cyclotron frequency in the regions where the flare energy is thought to be released. The goal is not only to identify all bursts in the meter and decimeter spectrogram or in HXR lightcurves, but patterns in the combined data set. Such patterns are needed for a fresh look at the decimeter burst emission processes. Although many decimeter burst emission mechanisms have been proposed (review by Benz, 2002), few are supported by further observations. Thus the investigation of decimeter flare emissions must be considered to be still in an exploratory phase. 

The ultimate goal of the comparison with HXR emission is to make coherent radio emissions useful diagnostic tools for the understanding of the flare processes. Thus the strategy of this survey is to start from HXRs as a proxy for energy release and particle acceleration. 

\section{Observations and Methods}
The Reuven Ramaty High Energy Solar Spectroscopic Imager (RHESSI) was launched into orbit on 5 February 2002 (Lin {\it et al.}, 2002).  The germanium detectors register photons in the energy range from 3 keV to 17 MeV with 1 keV resolution at low energies (Smith {\it et al.}, 2002). Imaging is achieved by nine absorbing grids modulating the observed flux by satellite rotation. For this work, the unprecedented spectral resolution at the lowest non-thermal energies was of primary importance. RHESSI observes the Sun full time, interrupted only by satellite night and the South Atlantic Anomaly (SAA). Energetic electrons sometimes cause spurious counts when the satellite is at high geomagnetic latitudes. We use standard RHESSI software (Schwartz et al., 2002) and calibrations updated to mid-March 2004. For the selection and analysis of a large number of flares, the HESSI Experimental Data Center (HEDC\footnote{HEDC is operated by the Institute of Astronomy at ETH Zurich and can be accessed via the URL http://www.hedc.ethz.ch/}, Saint-Hilaire et al., 2002) was used.

The Phoenix-2 spectrometer in Bleien (Switzerland) records continuously the solar radio emission from sunrise to sunset. The full Sun flux density and circular polarization are measured by the frequency-agile receiver in the range from 0.1-4.0 GHz at an integration time of 500 $\mu$s per measurement. A description of the instrument can be found in Benz et al. (1991) and Messmer, Benz, \& Monstein (1999). Since the launch of RHESSI, Phoenix-2 observes an overview from 100 MHz to 4.0 GHz in channels known to be low on terrestrial interference. The spectrum was sampled evenly with about an equal number of channels per GHz. The channel bandwidth varies between 1, 3, and 10 MHz depending on frequency, and the sampling time of each channel was fixed at 100 ms throughout the survey. The antenna is calibrated at noon every day. In the period studied here, the instrument has operated automatically without major interruptions and changes.

The broad total bandwidth of Phoenix-2 is ideal to search for associated emission and can easily be browsed from the one hour scale to subsecond resolution. The data allow easy detections of broadband emissions down to the lower limit at 100 MHz, but are not sensitive to narrowband events in meter waves, such as type I bursts. For this type we have complemented the data with the observations of the Nan\c{c}ay Radioheliograph published in Solar and Geophysical Data (SGD), providing both intensity and position at several frequencies.

\subsection{Selection}
Flares from the RHESSI flare list in HEDC in the period from February 13, 2002, until June 30, 2003, were selected for comparison with Phoenix observations. The selection criteria were: {\sl (i)} flares larger than GOES class C5.0 and {\sl (ii)} separated more than 10 minutes between peak times. Events within 10 minutes were considered as one flare. Events without modulation and no flare position in the list are usually non- solar and were excluded. In doubtful cases, when no radio emission was present or only a type I noise storm, GOES-8 data were used to confirm the reality of the flare. A total of 670 RHESSI flares were found to satisfy these constraints. 

From this set, the flares were selected during which Phoenix-2 was operating and observing the Sun. This second step reduced the total of the selected flares for direct comparison to 201. 

In addition, we have complemented the survey by two data sets including smaller flares to study in more depth the HXR-radio relation for homologous flares and for noise storms.

\subsection{Association and Correlation}
The term {\sl association} is used here in the sense of radiations occurring in the same flare. A radio burst is defined as associated if it occurs during enhanced flare hard X-ray emission (usually 12 - 25 keV was used, higher for the largest events) or within about 5 minutes before or after. For a burst rate typical for high solar activity, a chance association of about 5 events or 2\% must be expected. Even if they originate in the same active region, associated emissions are not necessarily directly related to a single acceleration process. A direct relation is more plausible if the light curves of two associated emissions evolve simultaneously. Then the term {\sl correlation} is used. The correlation was roughly assessed by comparing the peak of the radio emission to the X-ray flare phases: before, during rise phase, peak, decay, or after the HXR emission. In case of several major peaks, each was considered individually. RHESSI observations are often interrupted by satellite night and the SAA. If RHESSI observations were not complete enough to determine the flare phase, GOES-8 data and their time derivative were used instead as a proxy for HXRs. 

Radio emissions of the 201 joint events were searched visually and at high resolution in spectrograms containing the full data. The standard Phoenix software packages, Ragview and RAPP Viewer, were used. It may be surprising that a considerable fraction of the radio bursts thus found are not recorded in our own list published in Solar and Geophysical Data. The list is done on a routine bases with little attention for the weakest events. Weak radio emission is often not reported in the presence of a other, dominating type of emission or can easily be missed due to sporadic interference. In the course of the close inspection of radio emissions, the flares were classified into burst types and subtypes. If no radio emission was found, the daily summary was checked for antenna pointing and the SGD for type I emission. 

\section{Results}
Of all 201 joint flares, 4 showed clearly the characteristics of extended flares in HXRs: duration longer than 10 minutes and gradual temporal variations. Some others appeared to be clearly thermal. However, with the much enhanced resolution of RHESSI the mentioned HXR classes appear less distinct and the definitions need to be revised. Therefore, we finally studied them not separately and combined all flares in the following.

\begin{table}
\caption{The association of hard X-ray flares with various types and subtypes of radio emissions. The peak time of radio emission of each subtype has been assessed relative to the HXR emission: before (up to 5 minutes) the HXR emission, in the rise phase, about coincident with the peak time, in the decay phase, and up to 5 minutes after the HXR emission. Each subtype of radio emission has been counted once per flare and timed according to its main peak.}
\vskip10mm
\noindent
\begin{tabular}{|l|r|r|r|r|r|r|}
\noalign{\hrule}
Type, Subtype&before &rise &peak &decay &after&total \cr
\noalign{\hrule}
Type III&&&&&& \cr
\hskip1cm regular metric&7&23&18&16&2&66\cr
\hskip1cm narrowband dcim&3&28&40&13&3&87\cr
\hskip1cm broadband m-dcim&1&4&9&&&14\cr
\hskip1cm reversed drift dcim&2&21&21&9&2&55\cr
\hskip1cm slow drift, dcim&1&4&2&3&&10\cr
\noalign{\hrule}
Pulsations&&&&&& \cr
\hskip1cm quasi-periodic metric&&&1&5&1&7\cr
\hskip1cm irregular, dcim&2&23&41&19&&85\cr
\hskip1cm drifting up, dcim&2&13&20&8&&43\cr
\hskip1cm diffuse, broadband&1&5&12&7&&25\cr
\noalign{\hrule}
Diffuse Continuum&&&&&& \cr
\hskip1cm broadband&&1&8&&&9\cr
\hskip1cm drifting up or type II&&2&9&3&1&15\cr
\hskip1cm narrowband patch&&3&4&3&1&11\cr
\hskip1cm drifting down&&2&2&&&4\cr
\noalign{\hrule}
Narrowband Spikes&&&&&& \cr
\hskip1cm metric&&4&2&&&6\cr
\hskip1cm decimetric&&5&14&2&&21\cr
\hskip1cm type IV associated&&&&2&&2\cr
\noalign{\hrule}
Type IV&&&&&& \cr
\hskip1cm pulsating&&&1&1&&2\cr
\hskip1cm intermediate drift &&&&1&&1\cr
\hskip1cm parallel drifting bands&&&&&&0\cr
\hskip1cm sudden reductions&&1&&1&&2\cr
\noalign{\hrule}
HF Broadband (Gyrosynch.)&&&&&& \cr
\hskip1cm hf broadband  &&1&73&&&74\cr
\hskip1cm hf broadband  only&&&2&&&2\cr
\noalign{\hrule}
\end{tabular}
\end{table}

\subsection{Flare Associated Radio Emissions}
The statistical results on flare associated radio emissions are presented in Table I. The peak emission of every radio subtype has been assessed from Phoenix-2 spectrograms and compared to the HXR flare phase. The flare peak time was defined by the interval including half a minute before and after a major HXR peak. The total number of radio classifications (541) exceeds the number of flares (201), as a flare may be accompanied by more than one type or subtype of radio emission. 

Table I indicates that type III emission is the most common radio emission during HXR flares. Only a fraction, however, is in the form of the classical metric type III bursts, generally interpreted as the signature of electron beams propagating upward in the corona into interplanetary space. Most flare associated type III bursts occur at decimeter wavelengths. Often their duration is a few tenths of seconds; the bandwidth is a few hundred MHz. If the bandwidth was too small to measure the drift rate with the available time resolution, events have been classified as `narrowband'. Furthermore, there are type III-like bursts drifting extremely slowly. All type III subtypes occur preferentially in the peak or rise phase of HXR emission. 

Broadband continuum at high frequencies increasing in flux to the limiting frequency of Phoenix-2 at 4 GHz (and apparently beyond) is generally interpreted as produced by gyrosynchrotron emission of mildly relativistic electrons. It peaks around 10 GHz in impulsive flares and around 3 GHz in extended flares. Thus, the 4 GHz upper limit of the observed band makes the Phoenix-2 spectrometer rather insensitive to the detection of this emission. Furthermore, the incoherent gyrosynchrotron process is less efficient and thus generally weaker than coherent emission. In our procedure of classifying the emission, we have not enhanced the sensitivity by integration. The 5 $\sigma$ detection limit at 4 GHz is 10 sfu. Thus the relatively low number of high-frequency broadband events is interpreted as the result of limited sensitivity. The emission reached always its maximum at peak HXR time with only one exception. The detection rate of high-frequency continuum generally increases with flare class, but we note here that the correlation with X-ray flux has a wide scatter. The correlation is considerably better in the series of homologous flares presented below. Some flares showed only high-frequency continuum but no other radio emission. They are listed separately in Table I.

The three specifically decimetric types including pulsations, diffuse continuum and narrowband spikes also have a significant trend to occur early in the HXR flare phase. Some pulsations clearly start before. Diffuse continua do not increase in flux at 4 GHz and are too narrow in bandwidth to originate from gyrosynchrotron emission. They have the spectrum of pulsations but less temporal variation. Narrowband spikes were found a factor of two less frequently than reported by G\"udel et al. (1991). This may be the result of lower frequency resolution in these survey data. 

For the classification of a type IV burst, a duration longer than 10 minutes was required following to the rules of Solar and Geophysical Data. This makes type IV bursts relatively rare events mostly occurring after large flares.

\begin{table}
\caption{Patterns in the association of radio emissions with HXR flares.}
\vskip2mm
\noindent
\begin{tabular}{|l|r|}
\noalign{\hrule}
Pattern&number\cr
\noalign{\hrule}
Standard&130\cr
Type III$_m$-only&8\cr
Afterglow&19\cr
Noise storm&10\cr
No coherent radio emission&34\cr
\noalign{\hrule}
\end{tabular}
\end{table}

\subsection{Patterns in Radio and X-ray Emissions: the Standard Flare}
The goal of classifying burst types and subtypes of radio emission is to order the emissions into homogeneous groups of common physical origin. As coherent radio emission processes can occur only under certain circumstances or the radiation may be absorbed by the coronal plasma, the same basic flare process may be accompanied by some subtype of radio emission one time, but not at some other time. Nevertheless, there may be trends in the association of radio and HXR emission that indicate different types of flares. In this section the flares have been loosely ordered into 5 patterns of radio and X-ray association. This ordering appears to be obvious, but it is not compelling. Table II lists the number of flares in each pattern.

A `standard pattern' of radio emission associated with HXR flare prevails. It consists of type III bursts indicating propagating electron beams, a broadband continuum increasing at 4 GHz apparently due to gyrosynchrotron emission, and some decimetric pulsation, diffuse continuum, or narrowband spikes in between. Figures 1 - 3 show examples thereof. There are events with neither type III emission, nor decimetric pulsations, spikes or continuum. In the `standard pattern' we include also those cases where some of the coherent emissions are missing. 

`Standard' thus means an HXR flare with according coherent meter and decimeter emissions in the main phase, where `according'  allows for smaller flares being generally accompanied by less types of radio bursts and vice versa. The latter is known as `large flares syndrome' (Kahler, 1982) and must be avoided in flare classification. It was the main reason for the impossibility to further divide the standard pattern without falling back into radio subclasses. 

\subsubsection{Homologous Flares}
Although flares of the standard pattern are very individualistic, there are sometimes clear similarities in `homologous flares' originating at the same location (or active region). Similarity between homologous flares has been reported previously in H$\alpha$, soft and hard X-rays (e.g. Zirin, 1983; Machado, 1985; Farnik, Hudson, \& Watanabe, 1997). The occurrence of homologous flares can affect flare statistics and was obvious also in this survey. Examples of homologous flares are presented in Figs. 2 and 3.  Both flares are impulsive and follow the standard pattern. The two flares appear clearly more similar than two flares chosen from different active regions. Similar plasma parameters and magnetic field configurations thus may lead to more alike patterns of coherent radio emissions. Homologous flares thus may be used to study flares of different size or the evolution of an active region.

To study the properties of homologous flares in more detail, we have concentrated on active region NOAA 9830 that crossed the solar meridian on 20 February 2002 at 18 degrees south. The region was very prolific in small flares. To enlarge the sample, we have selected flares down to C1.0 jointly observed with Phoenix-2. The association with the region was determined from the RHESSI flare position or, if not available, from the H$\alpha$ position (Solar and Geophysical Data). The radio emissions are very simple and consist mostly of type III bursts. All type III bursts occurred at decimeter wavelengths, some reaching even beyond 3 GHz. The predominance of decimeter type III bursts seems to be a characteristic of this active region. Only the flare of 23 February 2002, 14:00 UT, shows some diffuse decimeter continuum in addition, which does not increase beyond 4 GHz. This flare is unusual in the series also by its large number of type III bursts but low HXR peak flux. All flares follow the `standard flare' pattern. 

\begin{table}[here]
\caption{Comparison of X-ray and radio emissions from flares in active region NOAA 9830. The GOES-8 peak flux in the soft channel (1.5 - 12 keV) is given in $\mu$Wm$^{-2}$ (GOES class C) background subtracted. It is a measure of the thermal X-ray emission. The RHESSI count rate refers to the total counts by the front detector in the 12-25 keV band minus background. The Phoenix-2 radio flux in the 3.5 - 4.0 GHz band is given in solar flux units [sfu].}
\vskip2mm
\noindent
\begin{tabular}{|l|l|c|r|r|c|l|}
\noalign{\hrule}
Date&GOES &GOES &RHESSI&Flux&Number of&Frequency\cr
2002&peak &SXR&HXR &3.5-4&type III &range of\cr
&time&peak&peak&GHz&bursts&type III\cr
\noalign{\hrule}
26 Feb &10:27&9.6&8541&252.8&4&1.1 - 2.0 GHz\cr
23 Feb &14:00&4.3&192&13.3&29&0.6 - 2.5 GHz\cr
27 Feb &13:02&4.3&1241&11.4&3&400-600 MHz\cr
25 Feb &10:41&3.5&986&5.0&6&0.7 - 1.0 GHz\cr
23 Feb &14:37&3.4&104&2.7&0&\cr
25 Feb &15:54&3.3&800&7.3&17&1.0 - 2.7 GHz\cr
01 Mar &09:59&3.2&283&2.0&0&\cr
24 Feb &15:40&2.4&573&3.5&3&3.0 - 3.5 GHz\cr
23 Feb &11:58&2.1&123&3.9&0&\cr
01 Mar &15:32&2.0&203&4.1&0&\cr
17 Feb &13:49&1.8&92&2.5&0&\cr
17 Feb &15:07&1.7&-&$<$2.0&0&\cr
\noalign{\hrule}
\end{tabular}
\end{table}

Table III summarizes the results of the analysis of 12 homologous flares in type III radio emission and X-rays. They form a homogenous set of observations with constant attenuator state (1) and no decimation. There is an obvious correlation between soft X-rays (GOES peak flux) and hard X-rays (RHESSI counts). After integration in time (10s) and frequency (0.5GHz), most flares show a high-frequency continuum, indicating gyrosynchrotron emission. The association is in good agreement with the linear correlation between hard X-rays and centimeter emission reported earlier (e.g. Kosugi, Dennis, \& Kai, 1988). A significant correlation can been found in Table III also between hard X-rays (RHESSI) and soft X-rays (GOES). 

Surprisingly, the correlation between hard X-rays and type III bursts is weaker, as demonstrated in the numerical evaluation: The cross-correlation coefficients between the hard X-ray flux (reference) and the high-frequency continuum, soft X-rays, and number of type III bursts amount to 0.99, 0.94, and -0.02, respectively. The correlation is statistically significant for high-frequency continuum and soft X-rays, but not for type III bursts. Nevertheless, the relation between X-rays and the number of type III bursts is not random as demonstrated by the following tests: The number of type III bursts of the 6 flares with the lowest GOES class add up to only 3, whereas in the other half they amount to 59. Furthermore, type III bursts in our sample are only associated in flares with class C2.4 and higher. Of course, if this threshold applies to active region 9830 at all, it obviously does not for others. 

\subsection{The `Type III$_m$-only' Pattern}
A few flares showing strong meter-wave type III bursts but small HXR flux seem to form a separate class of events. They have no other decimeter emissions, nor a high-frequency continuum suggestive of gyrosynchrotron emission. Such groups of type III bursts are sometimes associated with `metric spikes' around 300 MHz. According to previous work, metric spikes are not well associated with H$\alpha$ flares or HXR events (Benz, Csillaghy, \& Aschwanden, 1996). Meter-wave type III bursts are observed to occur at high coronal altitudes and have been found to be connected to interplanetary electron events (Benz et al., 2001; Maia et al., 2001). 

Figure 4 displays an example, where the type III emissions are very intense at decimeter waves. Some bursts start at a frequency as high as 1.5 GHz, but few continue below 300 MHz. Thus the frequency range of the type III emissions is shifted to higher frequencies than usual. This seems to be related to the rather uncommon association of `type III - only' with enhanced HXRs. 

The occasional correlation of some type III bursts with HXR peaks has been noted before. In Figure 4, the first peaks in the 12 - 25 keV flux coincide with two groups (A and B). At higher resolution than used for Fig. 4, the drift properties of type III bursts can be determined. They show reversed slope bursts for group A (drifting to higher density) and U-bursts for B. The last prominent group (C) contains a strong burst with reversed slope. However, some of the non-correlating groups also contain reversed slope bursts and a systematic trend is not apparent. A study of the cases with good HXR and type III correlation is in preparation (Arzner \& Benz, 2004).

According to Table II the observed number of `type III$_m$-only' events is small. Large groups of metric type III bursts are very common during times of high solar activity. Our selection starting from HXR flares does not pick them because meter-wave type III bursts are rarely associated with HXR emission (see Introduction). 

\subsection{Radio Afterglows of HXR Flares}
The `afterglow' pattern of decimeter radio emission occurs occasionally after large flares (M-class and higher). The radio emission is accompanied by little or no HXR emission and is classified as decimetric pulsation or type IV burst. In Table I, the afterglow pattern appears as emissions in the decay phase or after the flare. Figure 5 shows a combination of the standard pattern (diffuse continuum and metric type III bursts) during the peak phase of the HXR emission, having decimetric `afterglows'. The flare reached a GOES class of M5.1 at 11:12 UT, but had two phases as suggested by the two phases of HXR emission, both followed by narrowband spikes and pulsations at decimeter waves. They are both accompanied by weak HXR emissions.

\subsection{Noise Storm Associations}
Strong type I emission (noise storm) at meter wavelength was noted in 33 flares. During solar maximum, there is usually some noise storm going on, thus the above number depends on sensitivity and includes chance associations. If the X-ray emission was associated also with another radio emission, it was not considered as a noise storm pattern. For the flares with no other radio emission, the positions of the X-ray emission were compared with noise storms using data from the Nan\c{c}ay Radioheliograph. Association was defined by spatial agreement within half a photospheric radius (8 arcminutes). In 10 cases (5\%), the X-ray emission was mostly or only associated with a noise storm. 

Figure 6 shows such a case of `noise storm associated X-ray emission'. The HXR flare lasted more than 3 hours and had the characteristics of the extended type. The rise of the HXR emission is accompanied by a diffuse continuum starting at 1.5 GHz and exceeding 4 GHz. It does not increase at 4 GHz and was not reported above 4 GHz in SGD. The relatively narrow bandwidth speaks against gyrosynchrotron emission. The type I emission has two enhancements: one at the time of the start of the continuum, the other, more prominent, at its end. Both consist of a chain of type I bursts starting around 340 MHz, drifting slowly to lower frequency. A noise storm continuum between 200 - 350 MHz is noticeable after some change in the color table or by integration (not visible in Fig. 6). It peaks at 09:38 UT. Figure 7 shows the location of the associated X-ray emission integrated from 09:36 - 38 UT. It is not a footpoint emission in the chromosphere, but is located at a surprising 60" projected above the solar disk. The radio emission at 327 MHz is separated from the X-ray emission by 350". The two emissions obviously do not originate from the same source and may not be directly related.


\begin{figure}
\begin{center}
\leavevmode
\mbox{\hspace{-0.1cm}\epsfxsize=11cm
\epsffile{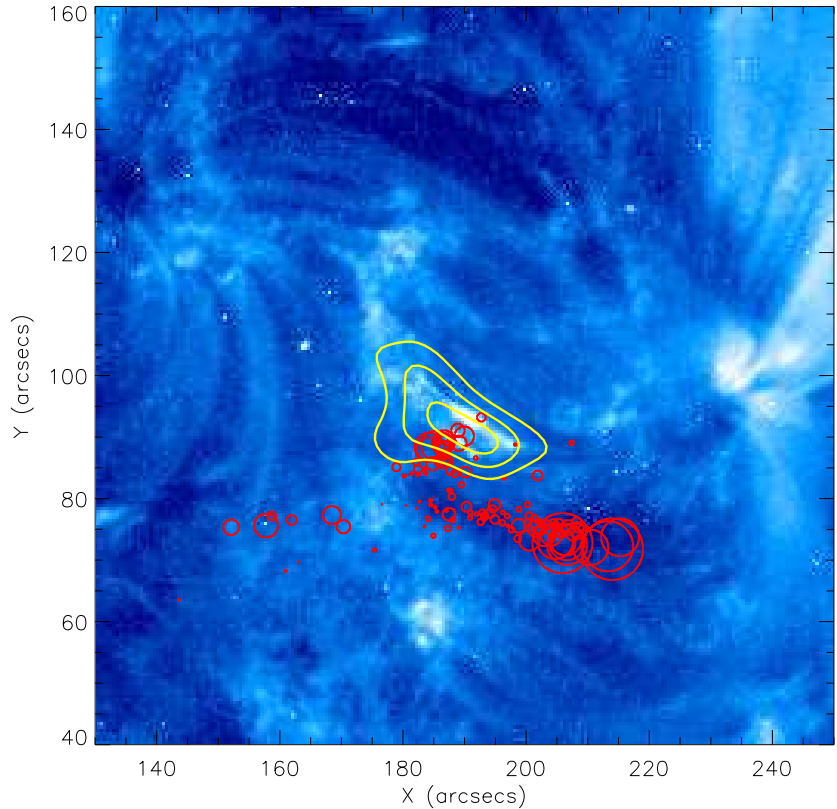}}
\end{center}
\vskip-0.2cm
\caption[]{A group of type I bursts has been observed with the VLA on 1 October 2002, 14:51 - 15:05 UT at 327 MHz. The centroid positions are given by circles whose radii indicate the intensity. RHESSI observations at 6 - 12 keV of a B9 flare at 15:01 UT are shown by isophotes. The observations are overlaid on a TRACE image taken at 15:29 UT in the 171 \AA\ band.}
\end{figure}

The relation of X-ray and type I emission has been further investigated in a joint RHESSI-VLA-Phoenix observation in another event shown in Fig. 8. The X-ray flare and the type I activity were considerably weaker than in the case of Figs. 6 and 7. The radio positions in Fig. 8 are scattered near the location of the X-ray emission and along a linear structure 13" to the south that appears dark at 171\AA\ (Fe IX and Fe X lines). During the X-ray flare, the type I emission originated mostly from sources within 5" (projected) from the RHESSI flare, but was not intense. The pronounced type I bursts occur without obvious relation to the X-ray light curve. The X-ray flare is located between two ribbons of 171\AA\ emission that lighten up during the flare and that are separated by only 4.2", thus suggesting small loop sizes. As the sources in Fig. 8 are close to the center of the disk, projection plays a more serious role than in Fig. 7. The results of this investigation are consistent with the conclusion that there may be a general relation between noise storm and impulsive X-ray emission in the evolution of an active region, but there is no direct causal link.

\subsection{Radio-quiet Flares}
In 34 flares out of the 201 selected, no coherent radio emission was associated between 100 and 4000 MHz. The non-detections have been confirmed by consulting other flare listings, both radio and X-rays. As type I noise storms are observed nearly all the time during maximum years of solar activity, the flare position was checked against the noise storm position of the day. Only if no spatially coincident noise storm is found, the pattern `radio-quiet' is used. Of course, higher sensitivity may reveal some weak radio emission, particularly at centimeter wavelengths where gyrosynchrotron radiation must be expected. This is evident in Table III where the observed flux was integrated in a broad frequency band and over 10 seconds to improve the sensitivity by more than an order of magnitude. Nearly every HXR flare can be detected at 3.5 GHz by this method, which is not surprising in view of the well-known correlation between HXRs and gyrosynchrotron emission. For this reason, we ignored the high-frequency continuum in classifying `radio-quiet' flares and included in this pattern two flares that had no trace of a radio emission except an intense high-frequency continuum. 

An example of non-detection of radio emission is presented in Fig. 9. The radio light curves show broadband variations of about 5 sfu amplitude. Such variations occur all the time and are caused by antenna pointing errors. Nevertheless they demonstrate that the Phoenix-2 spectrometer was observing the Sun and has not detected any radio emission exceeding that limit. Using the same integrations as in Table III, an upper limit of flare gyrosynchrotron emission of 3.2 sfu was established in this case. Thus the flare must have been extremely low on relativistic electrons. Note that the X-ray flare presented in Fig. 9 can be classified as `thermal' as it shows no emission beyond 50 keV despite its large GOES class. Not all radio-quiet flares were thermal, however, and this property needs to be further studied.

The ratio of non-detections, amounting to 16.9$\pm$2.9\%, appears close to the previously reported one by Simnett \& Benz (1986). The extension to smaller events in X-ray magnitude must have decreased the association rate. On the other hand, the enhanced sensitivity and search range of frequencies has increased the association rate. The two opposing effects have apparently cancelled out. 

The radio-quiet events were generally weaker in soft X-rays than the radio-associated ones. This was established by a comparison of their occurrence in weak and strong flares. The percentage of radio-quiet flares is for\\
C-class events (C5.0 - C9.9): \ \ \ \ 24.1 $\pm$ 4.7\%\\
M-class events ($>$ M1.0):  \ \ \ \ \ \ \ \ 9.9 $\pm$ 3.5\%\\
The difference suggests the existence of a threshold mechanism for coherent radio emission that is more often exceeded in large flares. 

\section{Discussion}
We summarize below the main results of the present study of radio emissions between 100 MHz and 4 GHz during 201 HXR flares:
\begin{itemize}
\item The most frequent coherent radio emission during X-ray flares are type III bursts (Table I). A majority of 72\% of them occur in the decimeter wavelength range. Less than 10\% of these continue to meter waves. Type III bursts occur preferentially in the rise and peak phase of flare HXRs. Thus they represent beams of electrons accelerated in the main flare phase.  

\item The timing of type III bursts with HXR peaks is sometimes excellent, but more often there is no peak-to-peak correlation (Fig.4). Obviously, electrons usually reach their target site without type III emission. On the other hand, type III emission is also observed before the start of HXR emission, indicating weak electron acceleration in the pre-flare phase.

\item In the decimeter range, the type III emission is about as abundant as the intrinsically decimetric types (pulsations, diffuse continuum and narrowband spikes) together. The occurrence of decimetric emissions has its maximum in the peak HXR phase.

\item Although associated, the type III and the other decimetric emissions usually do not correlate well in time with the HXRs. To capture this somewhat broad association, we have defined five patterns. Most common is the `standard pattern' embracing 65\% of all flares, where type III, some decimetric emission, or both are associated in the main flare phase. Half of the remaining flares are associated with a decimetric `afterglow', a noise storm, or just metric type III bursts. The other half has no coherent radio emission (Table II). The differences are larger than can be explained by radio absorption alone and suggest qualitative different flare types.

\item In 5\% of all cases only a noise storm was in progress in the same active region. We note however that the sources of HXRs and type I emission are usually separated by a hundred arcseconds or more, and that the temporal correlation is not obvious. Although the two emissions originate from the same active region, a direct causal link is questionable.

\item In 17 \% of the X-ray flares larger than C5.0 no coherent radio emission was found in the spectral range of the survey. The percentage reduces to 10 \% for flares larger than M1.0. In addition to flare size, the electron spectrum seems to influence also to the appearance of coherent radio emission (Fig. 9). The trend noted for `radio-quiet' and `noise storm associated' flares to have `thermal' HXR spectra needs to be further investigated.

\item Both the general pattern of radio and X-ray emission and the detailed associations of radio subtypes become markedly more homogeneous, but not identical for flares from the same active region (homologous flares, Figs. 2 and 3, Table III). 

\item There is a good temporal correlation of high-frequency continuum, generally interpreted as gyrosynchrotron radiation, with HXR flare phase. The distribution with flare phase is much more concentrated at peak HXR phase than for the various types of coherent emission. The detection rate below 4 GHz, however, correlates with the HXR flux not as well as may be expected. This contradicts reports of good peak flux correlation at higher frequencies, where the gyrosynchrotron emission is usually optically thin. In the homologous flares observed from the same region, however, the correlation between gyrosynchrotron emission below 4 GHz and X-rays is evident.
\end{itemize}

\begin{figure}
\begin{center}
\leavevmode
\mbox{\hspace{0.1cm}\epsfxsize=11.7cm
\epsffile{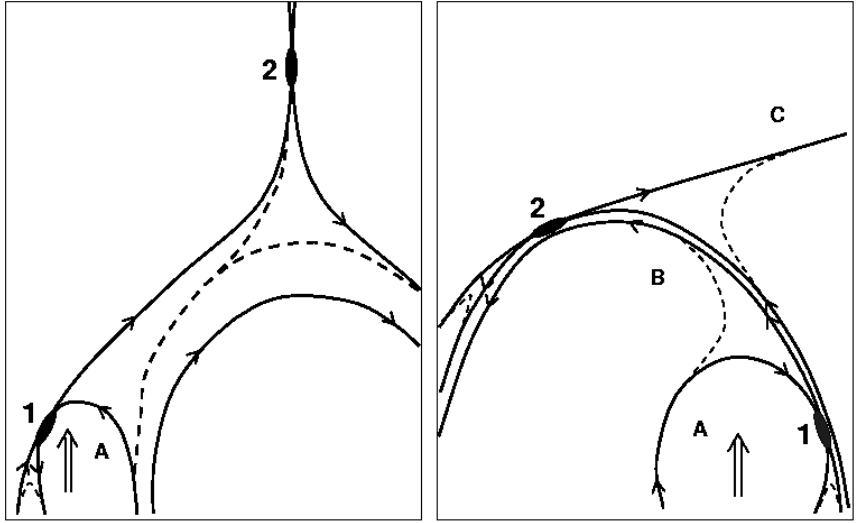}}
\end{center}
\caption[]{Schematic drawings illustrating the presence of two simultaneous reconnection sites (labeled 1 and 2) in a flare. Loop system A is the main driver, and reconnection 1 is the main energy release site. Reconnected field lines are indicated by dashed curves. {\it Left:} A helmet configuration is disturbed by the emerging loop system A. {\it Right:} Three magnetic field systems (A, B, and C) are involved. }
\end{figure}

Figure 10 shows possible scenarios involving two distant reconnection sites that may explain the loose association between X-ray and radio emissions of flares. The energy released at reconnection site 1 is larger, as the magnetic field is stronger and the driving is harder. Reconnection site 2 at high altitude may nevertheless become critical before reconnection 1 starts and will evolve differently. Reconnection 2 involves open field lines, allowing electron beams producing metric type III bursts to escape into interplanetary space. In both scenarios the reconnections are driven by a loop system A pushing up. Figure 10 Left depicts the situation of magnetic flux emerging into a helmet streamer configuration. Reconnection at point 1 intensifies reconnection at point 2 and vice versa. In Fig. 10 Right, the open field lines, C, are dragged out by the solar wind and bend over by solar rotation. They are related to the emerging A loops by the loop system B. 

Multiple reconnection sites may explain the relative independence of some flare associated radio emissions such as type III bursts and possibly some decimetric emissions. Reconnection sites at high altitude are suggested to be more prolific in coherent radio emissions, such as type III bursts, and reconnection sites at low altitude may dominate the total X-ray emission, but have radio beam signatures only in exceptional circumstances. We propose that in the 'standard pattern' the meter and most decimeter wave emissions originate near reconnection site 2, the X-rays near reconnection site 1. In the 'type III$_m$-only' and 'afterglows' patterns, reconnection may occur only at site 2, in the 'radio-quiet' pattern only at site 1. In 'noise storm' associated X-ray flares, the two reconnections sites seem to be without much physical connection. 

\section{Conclusions}
Hard and soft X-rays indicate the dominant energy release in flares. Radio emissions, on the other hand, have inspired flare physics for the past 50 years. In meter waves, electron beams and shock waves have been first discovered by their radio signatures. Centimeter waves have revealed the presence of relativistic electrons and are essential in present estimates of flare magnetic fields. At decimeter wavelengths, coherent radio emissions have been the starting point for various acceleration theories. Scenarios have emerged in the past decades where the coherent radio emissions fitted nicely into a standard flare model.

This comparison of X-rays with decimeter bursts has confirmed that the two emissions are relatively well associated, but surprisingly revealed them to be much more individualistic than simple scenarios would suggest. RHESSI resolves the low-energy spectrum, where small flares and flare elements can be more easily discovered and compared with radio emissions. Combined with the very broadband radio survey of Phoenix-2 in meter and decimeter wavelengths, the timing of two emissions was compared in a comprehensive way for the first time. Decimetric type III bursts and decimetric pulsations, spikes and continua are found preferentially in the peak hard X-ray phase, but often not simultaneously enough to suggest a common electron population or even a common acceleration process. The fact that HXRs and radio emissions do not correlate in detail suggests that there are other acceleration processes taking place in addition to the energetically dominant acceleration apparent in HXRs. Thus, the survey presented here indicates that the flare phenomenon is not a locally well-defined process but may include several parts of the active region yielding different HXR and radio signatures. 

If type III bursts occur, most are found in the rise or peak HXR phase. Their electron beams can be followed in the range of decimetric frequencies, suspected to include the plasma frequency of main flare energy release. Imaging type III emissions at various frequencies will make it possible to trace back these electron beams to their origin. Type III bursts thus have the potential to reveal energy release sites, secondary in amount but primary in time.

The other decimetric emission processes are still unknown. Nevertheless, some of the coherent flare radio emissions are suspected to be indicators for flare electron acceleration as they occur in the rise or peak phase of HXRs. Again, the temporal correlation is not sufficient to identify the radio and X-ray sources in a straightforward way. Therefore, the decimetric source regions do not seem to coincide always with the main flare energy release site nor to correlate with the acceleration process in a quantitative way. 

The decimeter flare emissions thus appear to be a largely unexplored field full of unexplained phenomena and unsolved issues of radiation processes. This comparative study has become possible only thanks to the recent developments in data retrieval and analysis tools. Further progress can be expected from new imaging instruments dedicated to solar observations, allowing to compare the radio sources with coronal structures and the flare geometry. Future instrumentation at higher sensitivity may possibly reveal the missing link between radio and X-ray signatures of non-thermal electrons.

\begin{acknowledgements}
Much of this work relied on the RHESSI Experimental Data Center (HEDC) supported by ETH Zurich (grant TH-W1/99-2). The construction of the Phoenix-2 spectrometer and the RHESSI work at ETH Zurich is partially supported by the Swiss National Science Foundation (grant nr. 20-67995.02). We thank the many people who have contributed to the successful operation of RHESSI and Phoenix-2 and to collecting this large set of data, in particular C. Monstein and H. Meyer and Dr. P. Messmer. We acknowledge the cooperation of Dr. T. S. Bastian of NRAO in Very Large Array observing and data analysis. The VLA is operated by Associated Universities, Inc. under contract with the US National Science Foundation. Data from the Nan\c{c}ay Radioheliograph have been used to locate some of the type I radio emission. \end{acknowledgements}


\end{article}
\end{document}